\documentclass{mem}
\usepackage{natbib}\usepackage{txfonts}\usepackage{balance}
\usepackage{graphicx}
\usepackage[a4paper,breaklinks,dvipdfm]{hyperref}
\idline{1}{1}
\begin{document}
\def\teff{$T\rm_{eff }$}
\def\kms{$\mathrm {km s}^{-1}$}

\title{
Young star cluster evolution and metallicity
}

   \subtitle{}

\author{
M. \,Mapelli\inst{1} 
\and A. \, Bressan\inst{2}
          }

%  \offprints{M. Mapelli}

\institute{
INAF --
Osservatorio Astronomico di Padova, Vicolo dell'Osservatorio 5,
I--35122 Padova, Italy, 
\email{michela.mapelli@oapd.inaf.it}
\and
Scuola Internazionale Superiore di Studi Avanzati (SISSA), Via Bonomea 265, 
I--34136 Trieste, Italy
}

\authorrunning{Mapelli \&{} Bressan}

\titlerunning{Young SCs and metallicity}

\abstract{
Young star clusters (SCs) are the cradle of stars and the site of important dynamical processes. We present $N-$body simulations of young SCs including recipes for metal-dependent stellar evolution and mass loss by stellar winds. We show that metallicity affects significantly the collapse and post-core collapse phase, provided that the core collapse timescale is of the same order of magnitude as the lifetime of massive stars. In particular, the reversal of core collapse is faster for metal-rich SCs, where stellar winds are stronger. As a consequence, the half-mass radius of metal-poor SCs expands more than that of metal-rich SCs.
}
\maketitle{}

\section{Introduction}
The evolution of a star cluster (SC) is connected with the metallicity of its stars, because the metallicity influences stellar winds (e.g. Kudritzki et al. 1987; Vink et al. 2001), remnant formation (e.g. Mapelli et al. 2009) and other properties of stars (e.g. Hurley et al. 2000). Young ($<100$ Myr) massive ($>10^3$ M$_\odot{}$) SCs may have a particularly short two-body relaxation time ($\approx{}10-100$ Myr, e.g. Portegies Zwart 2004). This implies that the core collapse and the post-core collapse phase occur on a timescale similar to the lifetime of massive stars. Thus, the peak of mass loss by stellar winds coincides approximately with the epoch of SC core collapse. The removal of mass by stellar winds and core-collapse supernovae (SNe) makes the SC potential well shallower, contributing to reverse the core collapse. In this paper, we discuss the results of $N-$body simulations of young SCs, including recipes for metal-dependent stellar evolution and stellar winds. We show that stellar metallicity can significantly affect the structural properties %(especially core radius, $r_{\rm c}$, and half-mass radius, $r_{\rm hm}$) 
of SCs in the early post-core collapse phase.

%\begin{figure}[]
%\resizebox{\hsize}{!}{\includegraphics[clip=true]{fig1.eps}}
%\caption{
%\footnotesize
%Mass loss rate for main sequence (MS) stars. Dotted green line: standard implementation in Starlab. Solid line: new implementation from Vink et al. (2001). Red line: 0.01 Z$_\odot{}$, black line: 0.3 Z$_\odot{}$, blue line: 1 Z$_\odot{}$.
%}
%\label{fig:fig1}
%\end{figure}

\begin{figure}[]
\resizebox{\hsize}{!}{\includegraphics[clip=true]{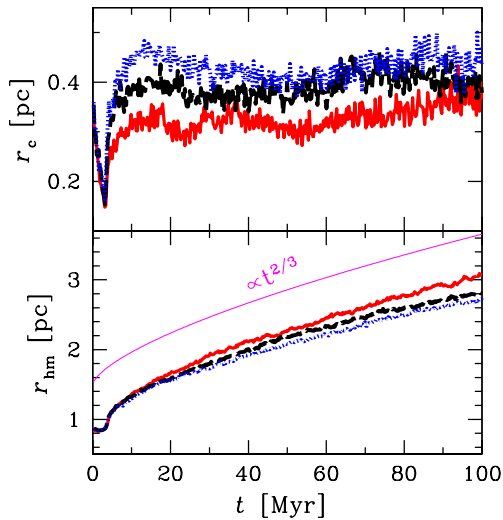}}
\caption{
\footnotesize
Top: core radius $r_{\rm c}$ as a function of time $t$. Bottom: half-mass radius $r_{\rm hm}$ as a function of $t$. Red solid line: metallicity $Z=0.01$ Z$_\odot{}$; black dashed line: 0.1 Z$_\odot{}$; blue dotted line: 1 Z$_\odot{}$. 
%Each line is the median of 100 runs. 
Magenta line in the bottom panel: analytical prediction ($r_{\rm hm}\propto{}t^{2/3}$). %In the insert: zoom of the first 7.5 Myr.
}
\label{fig:fig1}
\end{figure}

%\begin{figure}[]
%\resizebox{\hsize}{!}{\includegraphics[clip=true]{fig3.eps}}
%\caption{
%\footnotesize
%Half-mass radius as a function of time for metallicity $Z=0.01$ Z$_\odot{}$ (red solid line), 0.1 Z$_\odot{}$ (black dashed line), 1 Z$_\odot{}$ (blue dotted line). Magenta line: analytical prediction ($r_{\rm hm}\propto{}t^{2/3}$). In the insert: zoom of the first 7.5 Myr.
%}
%\label{fig:fig3}
%\end{figure}

\section{Simulations}
We perform $N-$body simulations of SCs using the Starlab public software environment (Portegies Zwart et al. 2001). We modified Starlab, to include metal-dependent stellar evolution and recipes for stellar winds by Vink et al. (2001; see also Mapelli et al. 2013 for more details on the code).
We simulated young intermediate-mass SCs, generated according to a multi-mass King model, with total mass $M_{\rm TOT}=3000-4000$ M$_\odot{}$, initial core radius $r_{\rm c}=0.4$ pc, concentration $c=1.03$. The stars in the SC follow a Kroupa (2001) initial mass function. We consider three different metallicities: $Z=0.01$, 0.1 and 1 Z$_\odot{}$. We ran 100 realizations of the same SC (by changing the random seed) for each metallicity, to filter out the statistical fluctuations.
\section{Results and conclusions}
The core collapse occurs at $t\sim{}3$ Myr in all simulations (top panel of Fig.~\ref{fig:fig1}). The reversal of core collapse depends strongly on the stellar metallicity, as the post-collapse $r_{\rm c}$ is a factor of $\approx{}1.5$ larger in metal-rich SCs than in metal-poor SCs. The reason is that mass loss by stellar winds and SNe is stronger in metal-rich SCs, making the core potential well shallower. At later times, the core radii at different $Z$ become similar again, as the effect of stellar winds is over, and the evolution of the core is completely determined by three-body encounters. The half-mass radius ($r_{\rm hm}$, bottom panel of Fig.~\ref{fig:fig1}) starts expanding after core collapse, according to the well-known analytical model ($r_{\rm hm}\propto{}t^{2/3}$, e.g. Elson et al. 1987). On the long-term evolution ($t\sim{}100$ Myr), the half-mass radius in metal poor SCs is $\approx{}10$ per cent larger than in metal-rich SCs. The reason is that, in metal-poor SCs, the reversal of core collapse is slower, implying higher core densities for a longer time. This means that the rate of dynamical interactions in the core of metal-poor SCs is higher, and more kinetic energy is pumped into the halo, increasing $r_{\rm hm}$. 
This result is in agreement with the recent simulations by Schulman et al. (2012). Other studies (Downing 2012; Sippel et al. 2012) do not find important differences in $r_{\rm hm}$, as they consider systems with much longer relaxation time. 
%, long if compared with the evolution of massive stars. 
Our results open interesting perspectives on the study of the dynamical evolution of young massive SCs.
%\begin{table*}
%\caption{Abundances for TO stars in M 92}
%\label{abun}
%\begin{center}
%\begin{tabular}{lccccccc}
%\hline
%\\
%Star \# & [Fe/H] & $\sigma$ & [Mg/Fe] & [Ca/Fe] & [Ti/Fe] & [Cr/Fe] & [Ba/Fe] \%\
%\hline
%\\
%18  &$ -2.63 $ & $0.22 $& $-0.02 $ &$+0.21$ & $+0.28$ & $-0.29 $ &$+0.11$ \\
%21  &$ -2.57 $ & $0.27 $& $-0.55 $ &$+0.17$ & $+0.44$ & $-0.22 $ &$-0.18$ \\
%34  &$ -2.58 $ & $0.24 $& $+0.06 $ &$+0.04$ & $+0.40$ & $   -  $ &$-0.05$ \\
%46  &$ -2.38 $ & $0.22 $& $-0.31 $ &$+0.22$ & $+0.17$ & $-0.27 $ &$-0.28$ \\
%60  &$ -2.54 $ & $0.30 $& $+0.12 $ &$+0.43$ & $+0.44$ & $-0.06 $ &$+0.39$ \\
%350 &$ -2.37 $ & $0.30 $& $-0.17 $ &$+0.37$ & $+0.34$ & $  -   $ &$  -  $ 
%\\
%\hline
%\end{tabular}
%\end{center}
%\end{table*}

\begin{acknowledgements}
 We thank the developers of Starlab, and especially  P. Hut, S. McMillan, J. Makino, and S. Portegies Zwart.  We acknowledge the CINECA Award N.  HP10CXB7O8 and HP10C894X7, 2011. MM acknowledges financial support from INAF through grant PRIN-2011-1. AB acknowledges financial support from MIUR through grant PRIN-2009-1.
\end{acknowledgements}

\bibliographystyle{aa}

\end{document}